\documentclass[]{spie}  

\usepackage{amsmath,amsfonts,amssymb}
\usepackage{graphicx}
\usepackage[colorlinks=true, allcolors=blue]{hyperref}

\title{Development of the observatory software for the SKAO}
\begin{document} 
\author[a]{Thaddeus Kenny}
\author[b]{Stewart J. Williams}
\author[c]{Viivi Pursiainen}
\author[d]{Elizabeth S. Bartlett}
\author[e]{Brendan McCollam}
\author[f]{Andrew D. Biggs}
\author[g]{Sean Ellis}
\author[h]{Rupert Lung}
\affil[a, b,c,d,e,f]{UK Astronomy Technology Centre, STFC, United Kingdom}
\affil[g,h]{CGI, United Kingdom}

\authorinfo{Further author information: thaddeus.kenny@stfc.ac.uk}

\pagestyle{empty} 
\setcounter{page}{301} 

\maketitle

\begin{abstract}
The Observatory Science Operations (OSO) subsystem of the SKAO consists of a range of complex tools which will be used to propose, design, schedule and execute observations. Bridging the gap between the science and telescope domains is the key responsibility of OSO, requiring considerations of usability, performance, availability and accessibility, amongst others. This paper describes the state of the observatory software as we approach construction milestones, how the applications meet these requirements using a modern technology architecture, and challenges so far. 
\end{abstract}


\section{INTRODUCTION}
\label{sec:intro}  

The SKAO is "one global observatory operating two telescopes on three sites." with the global headquarters situated at Jodrell Bank, UK; SKA-Mid in South Africa and the SKA-Low in Australia. OSO is the software subsystem responsible for the operation of this world-class observatory. It is an ecosystem of web-based tools that will be used by astronomers, telescope operators and SKAO staff through  Graphical User Interfaces (GUIs) and Command-Line Interfaces (CLIs). OSO can be viewed as a semi-automated workflow through the observing life-cycle, and provides the link between the user and the science data product. 

Commissioning activities begin this year (2024) with Array Assembly 0.5: 4 dishes for SKA-Mid and 4 stations for SKA-Low being used to demonstrate a working end-to-end system. Some OSO tools are being used during this commissioning, while the development of others is focused on the longer-term road map for full observatory operations. 

OSO is being developed by three software teams – two based in the UK and one based in India – following a SAFe process\cite{Klaassen20}. The teams collaborate closely and contribute to all the repositories, though due to the broad nature of OSO areas of expertise have developed for each team. 

\section{REQUIREMENTS}
\label{sec:requirements}

As would be expected to operate an observatory on the scale of the SKAO, there are a range of complex functional requirements identified during pre-construction. Alongside standard requirements such as the creation and validation of proposals and scheduling blocks, OSO needs to support SKAO functionality such as commensality (the shared, concurrent use of observatory resources between projects), concurrent execution on subarrays, and proposals using both SKAO telescopes. The tools need to present these concepts intuitively to capture the astronomer's intention, before translating this into the telescope control domain and propagating downstream. 

The OSO tools have a wide user base, including astronomers, proposal reviewers, SKAO staff and telescope operators. Alongside the functional requirements expected by the observatory, a range of non-functional quality attributes have also been identified\cite{Natarajan16}. While considerations of each attribute applies to all the tools, the relative importance varies. On the astronomer-facing side, users expect a modern interface that is responsive and performative. SKAO is committed to accessibility, which has been considered when designing the web UIs. The database and tools which interact with the telescope control system have a particular focus on robustness, availability and security. 

\section{ARCHITECTURE OVERVIEW}
\label{sec:architecture}

\begin{figure} [ht]
   \begin{center}
   \includegraphics[height=10cm]{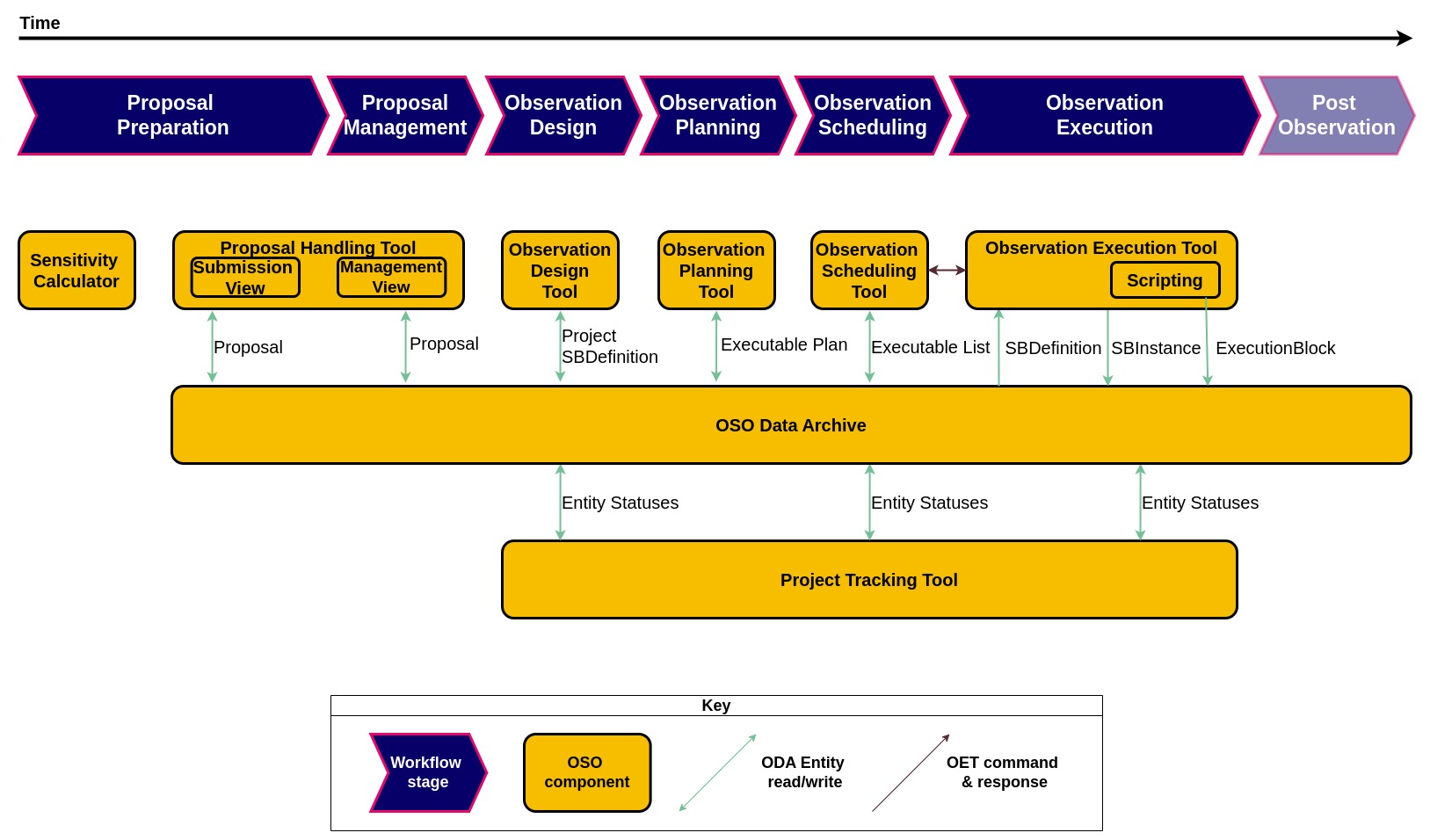}
   \end{center}
   \caption[example] 
   { \label{fig:workflow} OSO workflow, with applications and data flow to and from the ODA for each stage }
   \end{figure}

The observing life-cycle consists of several stages, from the preparation, submission, and assessment of observing proposals through to the design, scheduling, execution, and tracking of the resulting observations Each of these stages happen on different timescales relative to the observation being performed by the telescope, ranging from proposals  being prepared and submitted months in advance to the execution commands being sent in real time. Each stage is required to produce a data entity that is persisted for inspection and data linking, and the input to a stage requires the entity from the previous stage. For these reasons, OSO uses a Shared Data Architecture, where the responsibility for each stage is handled by an application which reads and writes the data to a central database. Figure \ref{fig:workflow} shows this workflow, with the OSO application responsible for each stage and the data entities they read or write. These applications and entities are explained in more detail in the following sections. 

\begin{figure} [ht]
   \begin{center}
   \begin{tabular}{c} 
   \includegraphics[height=10cm]{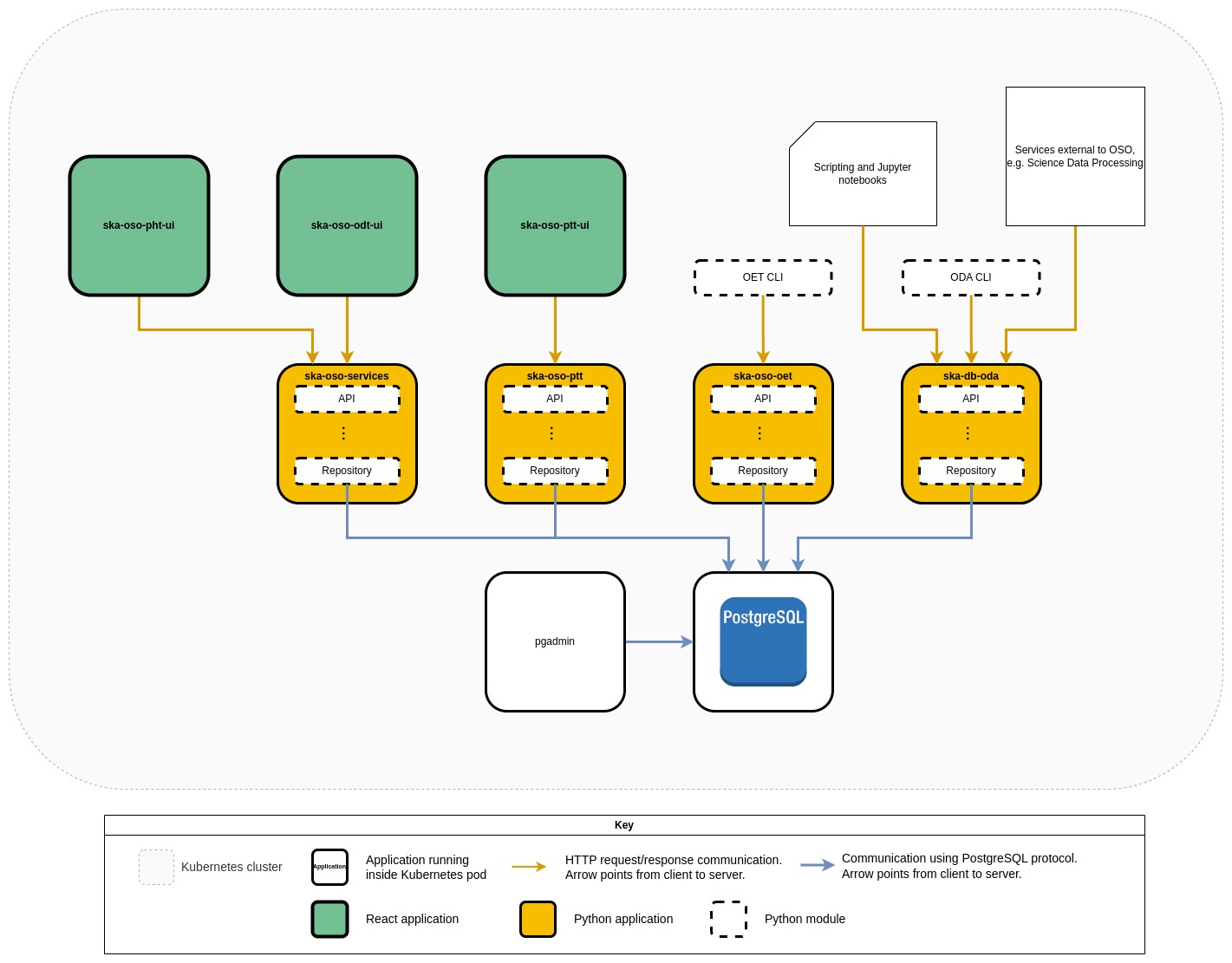}
   \end{tabular}
   \end{center}
   \caption[example] 
   { \label{fig:deployment} OSO deployment view as of 2024 }
   \end{figure}

In a shared data architecture, the services do not interact or communicate directly with each other. Instead, they access and transform data within the central data store – the OSO Data Archive (ODA). Due to the different time scales the applications operate on they pull the data when required, as opposed to an event driven architecture, where a stage will publish an event that is consumed by subsequent stages. Figure \ref{fig:deployment} shows a deployment view of the OSO architecture, with the interactions between each service, as it currently exists in the testing environment.  

A benefit of this architecture is that the services can all be built, tested and deployed in isolation. The dependency on the central database and model means the applications are not fully decoupled; however, as each application interacts with one or two of the data entities, they can evolve separately in most cases.

\section{TECHNOLOGY CHOICES}
\label{sec:technology}

OSO applications are developed using a modern technology stack and software engineering practices. The applications are mainly web based, with each application being containerised and deployed to Kubernetes. The applications expose their functionality as HTTP APIs, predominantly RESTful APIs. Users interact with OSO applications via GUIs and CLIs which interface with the low-level APIs, allowing control and presenting data in a more user-friendly format.

The GUIs are developed using Typescript and React, which was chosen because of its flexibility, modularity and popularity within the front-end community. The back-end services use Python, which is SKAO’s language of choice. All the RESTful APIs are defined using OpenAPI specifications, which the back-end requests and responses are validated against and the Typescript front end clients are generated from. This ensures a contract between the front and back ends. Each API service also uses the specification to offer a SwaggerUI, which is useful for testing and for stakeholders to exercise the APIs. 

SKAO has a central CI/CD (continuous integration and continuous delivery) infrastructure, which ensures every commit is built, tested, linted and documented. For OSO applications, each pipeline runs a set of integration tests where the application is tested with deployments of the lower-level dependencies and the shared data store. For more details on the CI/CD at SKAO see Di Carlo et al.\cite{DiCarlo22}

\section{DATA MODEL OVERVIEW}
\label{sec:pdm}

The SKA Project Data Model (PDM) is the reference implementation of the data models used and shared by OSO applications. It is released as a Python library and imported as a dependency in all the OSO applications. Pydantic\cite{Pydantic} is used to provide type safety and validation, plus serialisation and deserialisation. From the Pydantic models, we generate OpenAPI\cite{OpenAPI} schemas which can then be used in the API definitions for OSO applications that provide RESTful APIs. This ensures the PDM is the single source of truth for the OSO data models. 

The primary entities in the PDM are: 

\begin{itemize}
    \item \textbf{Proposal}: the astronomer domain description of an observation or set of observations on the telescope(s), often referred to as 'Phase 1'. It is designed and submitted by Investigators, contains all of the administrative details of a proposal (investigator details, proposal type, science category, science case, etc.) as well structured information about the observation (e.g. target information, spectral setup, etc.).
    \item \textbf{Project}: generated from an accepted Proposal, or in some cases may be created by SKAO staff without a related Proposal. Often referred to as 'Phase 2', it contains SBDefinitions grouped into Observing Blocks.
    \item \textbf{SBDefinition}: or Scheduling Block Definition, is the atomic unit of observing from the viewpoint of scheduling. It contains a set of instructions and configuration required to perform an observation. SBDefinitions are scheduled for execution, and can be cancelled, paused and ran more than once. SBDefinitions are mutable and can be edited by users and/or staff, e.g, to correct mistakes or tweak observing configurations.
    \item \textbf{SB Instance}: or Scheduling Block Instance, is a record of the specific version of an SBDefinition being sent for execution on the telescope, and contains any late binding arguments. Each execution of an SBDefinition results in an SBInstance. Typically, the SBInstance will contain a reference to a version of an SBDefinition, information about the execution environment such as the subarray and datetime, plus any extra arguments. An example would be a Target of Opportunity observation, where the target information is not known during SBDefinition creation.
    \item \textbf{ExecutionBlock}: is a record of the commands sent by OSO software to the telescope, and the responses received, during the execution of a SBInstance. The SBInstance can be thought of as what OSO requested from the telescope, and the ExecutionBlock as what actually happened during execution of the SBInstance. The ExecutionBlock serves another important purpose, in that its identifier is passed downstream to the data processing pipeline, which ultimately links the science data product back to the astronomer. This is also important for commissioning exercises, where an observation could be controlled by commands sent from a notebook, rather than the execution of a scheduling block. 
\end{itemize}

\begin{figure} [ht]
   \begin{center}
   \begin{tabular}{c} 
   \includegraphics[height=10cm]{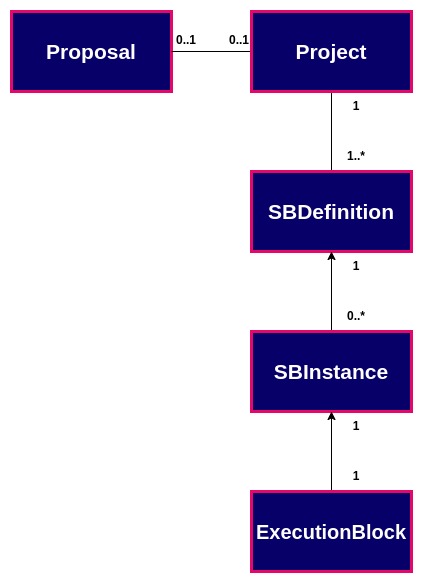}
   \end{tabular}
   \end{center}
   \caption[example] 
   { \label{fig:entities} OSO entities and their relationships }
   \end{figure}

There are simple relationships between these entities shown in figure \ref{fig:entities}, which map to the workflow shown in figure \ref{fig:workflow}.  

\section{ODA}
\label{sec:oda}

The OSO Data Archive (ODA) is the central data store in the Shared Data Architecture. It supports all the internal OSO processes and tools, persists all relevant data necessary to manage the proper behaviour of observation, and ultimately allows the science data products created by the SKA to be linked back to the astronomer.  

SKAO is a distributed system, with the two telescope sites, and HQ based at Jodrell Bank. OSO applications will be deployed at all three locations. The primary non-functional requirement for the ODA is availability: the reading of SBDefinitions and writing of SBInstances and ExecutionBlocks is critical to observing. The network connections at the remote sites cannot always be guaranteed so the ODA has been designed to guarantee availability over consistency, according to the CAP theorem, by being deployed at each site with a replication of the data. Each site will primarily be acting on different entities, and asynchronous replication will provide eventual consistency. Prototyping and development of this replication is ongoing, and during early commissioning activities the ODA deployments at the telescope sites will be independent.

The ODA consists of three main parts: 
\begin{enumerate}
    \item Repository and UnitOfWork abstractions which are imported into the OSO services that interact with the ODA. 
    \item A deployment of a PostgreSQL instance.  
    \item A Python application which offers a REST API to be used by clients external to OSO (e.g. the data processing pipeline).
\end{enumerate}

The Repository and UnitOfWork are concepts from Domain Driven Design\cite{Evans04}, and define an interface for OSO services to store and retrieve entities in the domain model. The ODA is designed as a generic database service which act on these entities; astronomy domain knowledge is delegated to the individual tools that interact with the ODA. The dependency of the OSO applications on these interfaces ensures the details of the specific database technology are abstracted away and could be changed in the future. A file-system implementation of the abstraction is also available, which demonstrates the abstraction is suitably generic and is used in some testing scenarios. 

The ODA has been under active development for the previous two years, and a version of the REST server and database are deployed and integrated at the telescope sites ahead of the first commissioning activities.  

Ahead of a final decision on ODA database technology, a working decision was made to use PostgreSQL as the database underpinning the ODA due to PostgreSQL's standing as an open source project, and its support for relational data and JSON types in the same data store.  At this stage of development for the telescope, the PDM and lower level model schema are not yet mature. As such, the decision was made to store most data in JSON format in the tables. Once the data models are more stable the data will be mapped to more relational database schema.

OSO applications that require fine-grained or direct access to the database can utilise the ODA code as a Python library, using the ODA's Repository and Unit Of Work implementations to interface directly with the database. The ODA Python application is implemented as one such application and is deployed alongside the database, offering a RESTful API for remote access of the ODA predominantly for non-OSO applications. Together the ODA library and ODA REST API ensure that a consistent approach is taken to AAA and interoperability.   

\section{PHT, ODT, PTT}
\label{sec:guis}

The Proposal Handling Tool (PHT), the Observation Design Tool (ODT) and the Project Tracking Tool (PTT) are the main GUIs developed by OSO. 

The PHT is responsible for creating and managing Proposals for the SKAO. It has a Submission Tool view which will be used by astronomers to propose an observation. After submission, the Handling Tool view will be used by reviewers and SKAO staff to assess and approve proposals for the cycle. The ODT will be used by operations staff to programmatically generate Projects and related SBDefinitions from Proposals, or to create Projects without a Proposal (e.g. projects created by SKA staff during commissioning activities). 

Both tools need to allow the user to configure an observation with the SKAO, translating between the astronomy domain and the telescope domain. Modern web users expect UIs to be intuitive, responsive and accessible. While it is expected that users will have a working knowledge of the astronomy behind the tools and have used similar tools for different observatories (e.g. the ALMA OT), SKAO will offer new functionality and configurations that need to be presented in an intuitive way. For example, elements of the data processing are scheduled in the same way as other telescope resources and configuration needs to be defined during observation design and sent with execution. 

The PTT will be used by SKAO staff to monitor Projects and the related entities, and perform updates to their statuses. While this is a different user base, many of the requirements are shared with the other services, such as the importance of role-based access control. The tools also need to be able to handle varying peak loads, which is expected to be achieved through horizontal scaling in the Kubernetes cluster. For example, towards the end of a proposal submission cycle it is expected the PHT will receive a large spike in submissions. 

All three tools have a similar architecture: React applications which communicate with a Python RESTful API, which in turn interacts with the ODA. The intention is that as much astronomy logic as possible is performed by the back-end services, with only basic validation happening in the browser to provide responsive feedback to the user. While the tools have different, overlapping user bases, it is expected they will have the same look and feel and many of the React components are being shared across the OSO teams working on these applications. There are also shared back-end API services for functionality such as target resolution. 

\begin{figure} [ht]
\begin{center}
\begin{tabular}{c} 
\includegraphics[height=8cm]{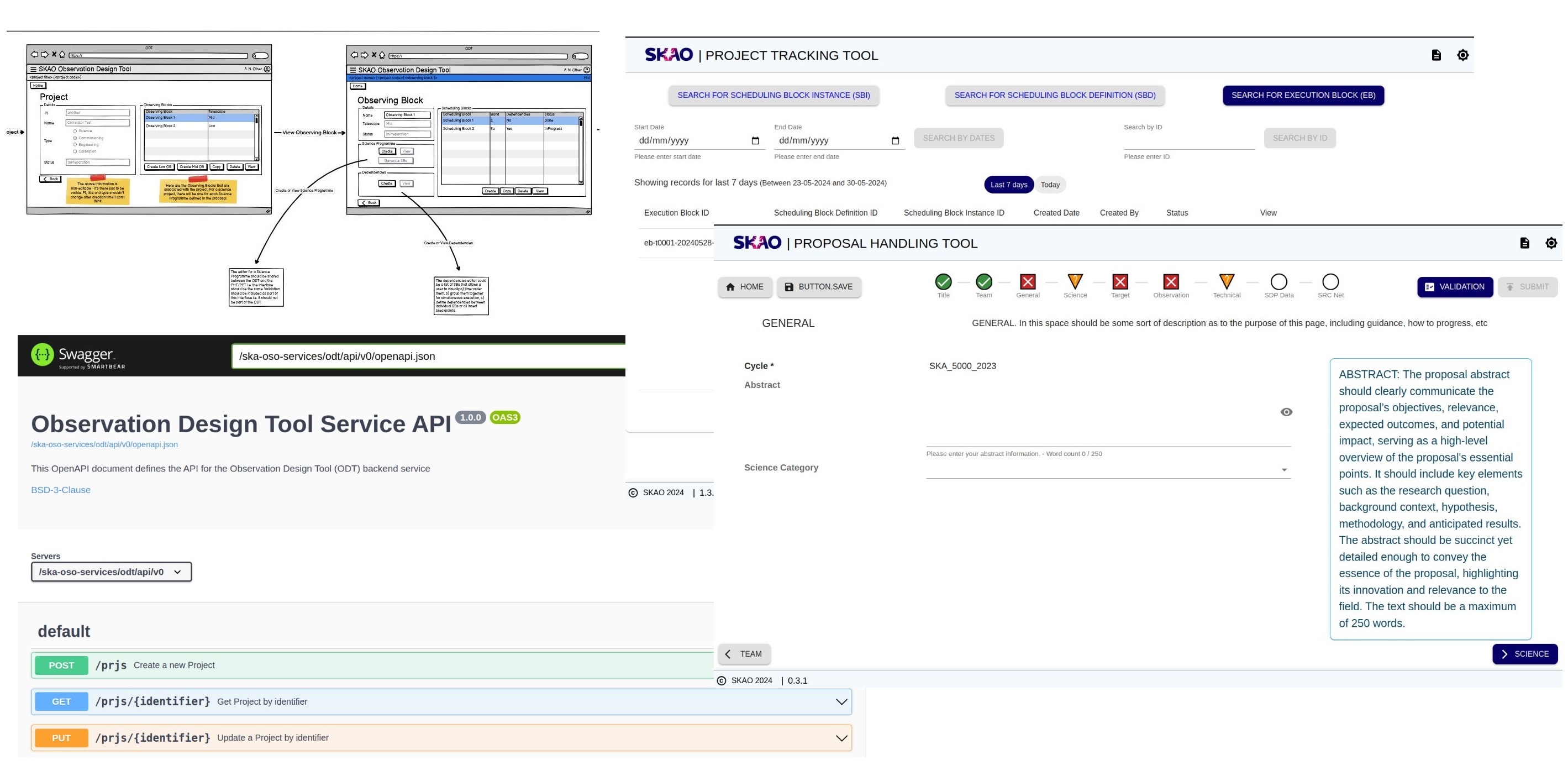}
\end{tabular}
\end{center}
\caption[example] 
{ \label{fig:guis} Screen shots of: initial wire-frames for the ODT; the SwaggerUI for the ODT API; initial versions of PTT and PHT }
\end{figure}

The tools are under active development and offer some core basic functionality in a test environment, including end-to-end interactions with the ODA. Initial wire-framing has been done with the input of science operation stakeholders and continues to be iterated on. This involves using online tools to create simple, non-functional views of the UI to demonstrate the user experience. These views are then being used by the developers during the implementation of the the UIs, which is currently focused on delivering functionality rather than the complete look and feel. Screenshots of this wire-framing and implementation can be seein in figure \ref{fig:guis}.

\section{PLANNING AND SCHEDULING}
\label{sec:planning}

The Observation Planning Tool (OPT) running at SKAO HQ creates a prioritised list of SBDefinitions spanning several days of observing, assessing factors such as proposal grade, fair return, and project completeness. The updated list is regularly pushed to the remote telescope sites where the Observation Scheduling Tool (OST) assess the prioritised list in the context of current and near-term observing conditions and resource availability. The OST selects the best observation(s) to perform in that moment and instructing the Observation Execution Tool (OET) to run the identified SBDefinition. The OST and OET are the only tools within the OSO architecture that interact directly with each other: the OST will send commands to the OET and the OET will send back status updates.  

Development on the OST and OPT is scheduled to start in 2025 as they are not critical for initial operation of the observatory. During commissioning, operators will choose which SBDefinitions to execute and send those commands to the OET using a manual interface. 

During full operations, the OSO workflow the expected to be mostly automated: an accepted Proposal will be used to generate a Project and SBDefinitions, which will be then planned, scheduled and sent for execution. A 'human-in-the-loop' will perform some actions and quality checks.

\section{OET AND SCRIPTING}
\label{sec:oet}

The Observation Execution Tool (OET) is responsible for executing observations on the two SKAO telescopes. Separate instances of the robust, multiprocessing script execution engine will provide telescope operators at each site an interface to start, pause, abort and monitor SBDefinition execution. 

Scheduled SBDefinitions will be sent for execution alongside and late binding arguments, either by the OST or through manual commands. The OET will create the SBInstance, fetch the related script and then run this script with the configuration data from the scheduling block. The logic which builds the telescope control commands from the SBInstance information is delegated to the scripts that are executed. Functionality has been built to allow the scripts to be updated via Git without having to rebuild and redeploy the OET, which will be important during commissioning activities. 

The OET is discussed in more detail in Pursiainen et al. (these proceedings).

\section{SENSITIVITY CALCULATOR}
\label{sec:senscalc}

The final tool developed by OSO is the Sensitivity Calculator, which is now publicly available\cite{SensCalc}. Given an observation set up, it will provide the astronomers with a sensitivity or integration time that the observation will achieve or require. This will inform the proposal design and submission. It is accessed through a publicly available GUI and provides an API that the PHT will make use of. This is a functional, stateless tool which does not interact with the ODA. See Biggs et al. (these proceedings) for more information on this tool.

For the OSO teams, it has been a fruitful lesson in building modern astronomy applications on the web, in particular in the design of modern Javascript framework applications and the importance of the division of responsibilities between the UI and the services. The introduction of a OpenAPI specification has ensured the interface to the back-end is clearly defined for both developers and stakeholders and defines a contract for the front-end that a Typescript client has been generated from. This approach has then been followed when implementing the other tools.

\section{CHALLENGES}
\label{sec:challenges}

As with any software project, especially one with the size, complexity and timescale of the SKAO, there are challenges and opportunities for learning along the way. There is a wide range of technical knowledge required with OSO: modern front-end web development, Python applications, database management, knowledge of the telescope and TANGO control system, Kubernetes, plus a large amount of astronomy domain knowledge. Iteratively developing the OSO suite of applications in parallel and in an agile fashion has often leads to excessive context switching and a limited time to develop a comprehensive solution in any one area, with the accompanying risk that some challenges to the design have yet to be uncovered.

As the lower-level telescope control systems are also under development, the interfaces and methods of interaction change frequently. As OSO is the interface used to control the telescope, the new lower-level interfaces and the transformation of the PDM model into these telescope commands need to be supported, which can take away effort from the development of the core OSO tools. 

The Shared Data Architecture means all of the services are coupled to the ODA and the PDM, and though the services are decoupled in terms of the entities they interact with it still requires updates to the PDM to be propagated through the whole system and coordinated between the teams. When the telescope is operational, there will also be the added constraint that previous data must be accessible, and efforts are being made to ensure backwards compatibility. 
 
\section{Conclusion}

This paper has presented the design of the observatory software and details of the implementation so far as the SKAO approaches its first commissioning activities. The Shared Data Architecture consists of several tools responsible for different stages in the observatory workflow interacting with the central ODA. Core functionality of these tools is demonstrated with end-to-end ODA interactions in testing environments. A world-class observatory such as the SKAO requires world-class software, and the user experience and further complex functionality included in this is on track to be delivered in the coming years.

\bibliography{report} 

\begin{thebibliography}{1}

\bibitem{Klaassen20}
Klaassen, P.~D., WIlliams, S.~J., Nicol, M., Alberti, V., Bridger, A., Chrysostomou, A., Valame, S., Bartlett, E.~S., Canzari, M., Deolalikar, A., Lightfoot, J., McDermott, A., Pursiainen, V., Ribero, H., and Sabater, J., ``{Observatory science operations tool development for the SKA within a scaled agile framework},'' in [{\em Observatory Operations: Strategies, Processes, and Systems VIII}{\nolinebreak\hspace{0.1em}]},  Adler, D.~S., Seaman, R.~L., and Benn, C.~R., eds.,  {\bf 11449},  114490Z, International Society for Optics and Photonics, SPIE (2020).

\bibitem{Natarajan16}
Natarajan, S., Barbosa, D., Barraca, J.~P., Bridger, A., Choudhury, S.~R., DiCarloe, M., Dolci, M., Gupta, Y., Guzman, J., den Heever, L.~V., Gerhard~LeRoux, M.~N., Patil, M., Smareglia, R., Swart, P., Thompson, R., Vrcic, S., and Williams, S., ``{SKA Telescope Manager (TM): Status and Architecture Overview},'' in [{\em Software and Cyberinfrastructure for Astronomy IV}{\nolinebreak\hspace{0.1em}]},  Chiozzi, G. and Guzman, J.~C., eds.,  {\bf 9913},  991302, International Society for Optics and Photonics, SPIE (2016).

\bibitem{DiCarlo22}
Carlo, M.~D., Harding, P., Yilmaz, U., Maia, D., Ribeiro, B., Nunes, D., Regateiro, D., Morgado, J., Paulo, M., Santos, M., Marotta, G., and Dolci, M., ``{CI-CD practices at SKA},'' in [{\em Software and Cyberinfrastructure for Astronomy VII}{\nolinebreak\hspace{0.1em}]},  Ibsen, J. and Chiozzi, G., eds.,  {\bf 12189},  1218903, International Society for Optics and Photonics, SPIE (2022).

\bibitem{Pydantic}
``{Pydantic}.'' \url{https://docs.pydantic.dev/latest/}.
\newblock Accessed: 2024-06-01.

\bibitem{OpenAPI}
``{OpenAPI specification}.'' \url{https://spec.openapis.org/oas/latest.html}.
\newblock Accessed: 2023-06-01.

\bibitem{Evans04}
Evans, E.,  [{\em Domain-Driven Design: Tackling Complexity in the Heart of Software}{\nolinebreak\hspace{0.1em}]}, Addison-Wesley (2004).

\bibitem{SensCalc}
``{SKAO Sensitivity Calculator}.'' \url{https://sensitivity-calculator.skao.int/}.
\newblock Accessed: 2024-06-01.

\end{thebibliography}
\bibliographystyle{spiebib} 

\end{document}